\documentclass[superscriptaddress,aps,pra,twocolumn,showpacs,nofootinbib,longbibliography]{revtex4-1}
\usepackage{etex}
\usepackage{amsmath,amssymb,amsthm}
\usepackage[colorlinks=true,citecolor=blue,urlcolor=blue]{hyperref}
\usepackage[pdftex]{graphicx}
\usepackage{times,txfonts}
\usepackage{braket}
\usepackage{color}
\usepackage{natbib}
\usepackage{float}
\usepackage{tabularx, booktabs}
\usepackage{amssymb}
\usepackage{graphics,epstopdf}
\usepackage{graphicx}
\usepackage{amsfonts}
\usepackage{caption}
\usepackage{subcaption}
\usepackage{wrapfig, framed, caption}
    \usepackage{lipsum}

\newcommand{\be}{\begin{equation}}
\newcommand{\ee}{\end{equation}}
\newcommand{\ba}{\begin{eqnarray}}
\newcommand{\ea}{\end{eqnarray}}

\newcommand{\tr}{\operatorname{Tr}}

\newtheorem{conjecture}{Conjecture}

\newtheorem{scenario}{Scenario}

\begin{document}

\title{Facets of bipartite nonlocality sharing by multiple observers via sequential measurements}

\author{Debarshi Das}
\email{debarshidas@jcbose.ac.in}
\affiliation{Centre for Astroparticle Physics and Space Science (CAPSS),
Bose Institute, Block EN, Sector V, Salt Lake, Kolkata 700 091, India}

\author{Arkaprabha Ghosal}
\email{a.ghosal1993@gmail.com}
\affiliation{Centre for Astroparticle Physics and Space Science (CAPSS),
Bose Institute, Block EN, Sector V, Salt Lake, Kolkata 700 091, India}

\author{Souradeep Sasmal}
\email{souradeep@mail.jcbose.ac.in}
\affiliation{Centre for Astroparticle Physics and Space Science (CAPSS),
Bose Institute, Block EN, Sector V, Salt Lake, Kolkata 700 091, India}

\author{Shiladitya Mal}
\email{shiladitya.27@gmail.com}
\affiliation{Harish-Chandra Research Institute, HBNI, Chhatnag Road, Jhunsi, Allahabad 211 019, India}

\author{A. S. Majumdar}
\email{archan@bose.res.in}
\affiliation{S. N. Bose National Centre for Basic Sciences, Block JD, Sector III, Salt Lake, Kolkata 700 098, India}

\begin{abstract}
Recently, it has been shown that at most two observers (Bobs) can sequentially demonstrate bipartite nonlocality with a spatially separated single observer (Alice) invoking a scenario where an entangled system of two spin-$\frac{1}{2}$ particles are shared between a single Alice in one wing and several Bobs on the other wing, who act sequentially and independently of each other [\href{https://journals.aps.org/prl/abstract/10.1103/PhysRevLett.114.250401}{Phys. Rev. Lett. {\bf 114}, 250401 (2015)}]. This has been probed through the quantum violations of CHSH inequality, i. e., when each observer performs two dichotomic measurements. In the present study we investigate how many Bobs can sequentially demonstrate bipartite nonlocality with a  single Alice in the above scenario when the number of measurement settings per observer is increased. It is shown that at most two Bobs can exhibit bipartite nonlocality with a single Alice using local realist inequalities with three as well as four dichotomic measurements per observer. We then conjecture that the above feature remains unchanged contingent upon using local realist inequalities with $n$ dichotomic measurements per observer, where $n$ is arbitrary. We further present the robustness of bipartite nonlocality sharing in the above scenario against the entanglement and mixedness of the shared state. 
\end{abstract}

\maketitle

\section{INTRODUCTION} 
Ideas and notions involved in classical mechanics significantly differ from the fundamental tenets of quantum mechanics. A seminal concept showing incompatibility between classical and quantum mechanics is quantum nonlocality. Quantum nonlocality was first pointed out by Einstein, Podolsky and Rosen (EPR) \cite{epr} and it has been intensively studied since the pioneering discovery by John Bell in 1964, known as Bell's theorem \cite{Bell}. Bell's theorem, in the form of  an inequality, presents an experimentally testable criterion to  check whether a theory is incompatible with the assumptions of local realism. It states that local measurements on a spatially separated quantum system can lead to a set of correlations which cannot be explained by shared classical randomness or local hidden variable (LHV) models. This feature of quantum correlations is called quantum nonlocality \cite{nonlocality}. Bell's inequality was later modified  by Clauser, Horne, Shimony and Holt (CHSH) \cite{CHSH}. Quantum nonlocality of a system certifies entanglement \cite{ent}. Realizing quantum violations of Bell-CHSH inequality in various quantum systems has acquired great interest as evidenced by a wide range of studies \cite{exp11,exp12,exp3,exp4,exp5,exp6,exp7,exp8,exp9}. Strong loophole-free Bell tests have also been reported \cite{lfexp1,lfexp2,lfexp3}.

An intriguing property of quantum correlations is that they are monogamous in nature. The restriction on sharing quantum correlations between several numbers of spatially separated observers is quantitatively expressed through the monogamy relations for entanglement \cite{monen} or the monogamy relations for Bell-nonlocality \cite{monnon}. In case of the multipartite scenario considered in the monogamy context, $n$ observers share $n$ number of particles, one particle per each observer, and all the observers are spatially separated. Hence, the no-signalling condition (the probability of obtaining one party's outcome does not depend on spatially separated other party's setting) is satisfied between any pair of observers. 

Recently, it has been shown \cite{sygp, mal}  that monogamy relations no longer hold for the correlations obtained from measurements performed by different parties if a different scenario is considered where the no-signalling condition is relaxed between a subset of observers. Here, relaxing the no-signalling condition does not imply violating relativistic causality, rather it implies a scenario where sequential measurements are performed by different observers on the same particle. Specifically, the scenario is that one observer (say, Alice) has access to one half of an entangled system of two spin-$\frac{1}{2}$ particles, whereas, multiple observers (say, multiple Bobs) can access and measure on another half of that entangled system sequentially. Here Alice is spatially separated from multiple Bobs. In this scenario it has been shown  \cite{sygp, mal} that at most two Bobs can demonstrate nonlocality with single Alice when each Bob performs different measurements with equal probability and when the measurements of each Bob are independent of the choices of measurement settings and outcomes of the previous Bobs. This result has also been confirmed by experiment \cite{exp1, exp2}. Note that the above result is probed through the quantum violations of CHSH inequality \cite{CHSH}, i. e., in the scenario where each observer performs two dichotomic measurements.

Sharing entanglement by multiple Bobs in the above scenario has been demonstrated recently \cite{bera}. Steerability of local quantum coherence by multiple observers has been studied recently \cite{saunak}. Sharing of tripartite nonlocality by multiple sequential observers has also been studied \cite{sahanew}. Quantum steering \cite{steer1, steer2, steer3} of a single system sequentially by multiple Bobs in the above scenario has also been investigated \cite{sas}, going beyond the monogamy restriction on steering \cite{monsteer1, monsteer2}. Moreover, it has been shown that the maximum number of Bobs who can sequentially steer single Alice increases with increasing number of measurement settings performed by each party \cite{sas}. More recently, it has been demonstrated that the number of Bobs who can steer the state of Alice increases with increase in $d$ when Alice and multiple Bobs share symmetric entangled state of arbitrary dimension $d \times d$ \cite{shenoy}.
 
In the present study we investigate how many Bobs can sequentially demonstrate bipartite nonlocality with single Alice in the above scenario when the number of measurement settings performed by each party is increased. Unlike steering, the interesting result revealed by the present study is that at most two Bobs can sequentially demonstrate bipartite nonlocality with single Alice when different local realist inequalities \cite{chain1, chain2, chain3, gisin, i3322, deng, bru, avis, gisin_open} with three or four dichotomic measurements per observer are used. In other words, no advantage over CHSH inequality is gained in the context of sharing of bipartite nonlocality when different local realist inequalities  \cite{chain1, chain2, chain3, gisin, i3322, deng, bru, avis, gisin_open} with three or four dichotomic measurements per observer are used. Based on this result we also conjecture that the above feature remains unchanged when each party performs $n$ dichotomic measurements.

One important point to be stressed here is that sharing of bipartite nonlocality \cite{sygp, mal} or steering \cite{sas} has been probed assuming that Alice and multiple Bobs initially share pure maximally entangled state. However, in real practical scenario it is very difficult to prepare pure maximally entangled state. Hence, in order to incorporate inaccuracies that appear in real scenario, we also investigate sharing of bipartite nonlocality by multiple Bobs with single Alice, when Alice and multiple Bobs initially share non-maximally entangled pure state or mixed state. The robustness of different local realist inequalities in the context of sharing of nonlocality against entanglement and mixedness of the shared state is explored in the present study. This investigation presents a new dimension in demonstrating inequivalence between different local realist inequalities with respect to sharing of nonlocality. Interestingly, we show that CHSH inequality is the most robust against entanglement and mixedness of the shared state in the context of sharing of nonlocality.

We organize this paper in the following way. In Sec. \ref{sec2}, we discuss in brief the concept of bipartite nonlocality in the context of the local realist inequalities that we use later. Next, in Sec. \ref{sec3}, we discuss the scenario in which we demonstrate bipartite nonlocality sharing. Unsharp measurement formalism used by us is also discussed in Sec. \ref{sec3}. Sharing of bipartite nonlocality by multiple Bobs with single Alice using local realist inequalities having three and four dichotomic measurements per observer is demonstrated in Sec. \ref{sec4} and Sec. \ref{sec5}, respectively. In Sec. \ref{sec6} we present the issue of bipartite nonlocality sharing by multiple Bobs with single Alice when Alice and multiple Bobs initially share non-maximally entangled pure states as well as mixed states. Finally, we end with concluding remarks in Sec. \ref{sec8}.

\section{Recapitulating bipartite nonlocality}\label{sec2}

Suppose $A \in \mathbb{F}_{\alpha} $ and $B \in \mathbb{F}_{\beta}$ are the possible choices of measurements for two spatially separated observers, say Alice and Bob, with outcomes $a \in \mathbb{G}_{a}$ and $b \in \mathbb{G}_{b}$, respectively. $A_1$, $A_2$ $A_3$, ..., $A_n$ denote the possible choices of measurement settings for Alice; $B_1$, $B_2$, $B_3$, ..., $B_n$ denote the possible choices of measurement settings for Bob. The joint probability of obtaining the outcomes $a$ and $b$, when measurements $A$ and $B$ are performed by Alice and Bob locally on state $\rho^{AB}$, respectively, is given by, $P(a,b|A,B,\rho_{AB})$. The bipartite state $\rho_{AB}$ of the system is nonlocal \cite{Bell} iff it is not the case that for all  $A \in \mathbb{F}_{\alpha} $, $B \in \mathbb{F}_{\beta}$, $a \in \mathbb{G}_{a}$, $b \in \mathbb{G}_{b}$, the joint probability distribution can be written in the form
\begin{equation}
P(a, b|A, B, \rho_{AB}) = \sum_{\lambda} p_{\lambda} P_{\lambda}(a|A)P_{\lambda}(b|B)
\end{equation}
where $p_{\lambda}$ is the probability distribution over the hidden variables $\lambda$; $\sum_{\lambda} p_{\lambda} = 1$; $P_{\lambda}(a|A)$ and $P_{\lambda}(b|B)$ denote arbitrary probability distributions conditioned upon $\lambda$. In the present study we restrict ourselves to the experimental scenario involving only dichotomic measurements, i. e., measurements having two outcomes.

In $2 \times 2 \times 2$ experimental scenario (involving two parties, two measurement settings per party, two outcomes per setting), the necessary and sufficient criterion to detect nonlocality between Alice and Bob is given by \cite{CHSH},
\begin{equation}
C^2=|\langle A_1 B_1 \rangle + \langle A_2 B_1 \rangle + \langle A_1 B_2 \rangle  - \langle A_2 B_2 \rangle| \leq 2.
\label{chsh}
\end{equation}
where $A_1$ and $A_2$ denote the two possible choices of measurement settings for Alice; $B_1$ and $B_2$ denote the two possible choices of measurement settings for Bob. This inequality is known as the CHSH (Clauser-Horne-Shimony-Holt) inequality \cite{CHSH}.

For arbitrary number of dichotomic measurements performed by each party, quantum violation of Chained Bell-CHSH inequality \cite{chain1, chain2, chain3} demonstrates bipartite nonlocality between Alice and Bob. Chained $3$-settings Bell-CHSH inequality \cite{chain1, chain2, chain3} has the following form,
\begin{align}
Chain^3 &= |\langle A_1 B_1 \rangle + \langle A_2 B_1 \rangle + \langle A_2 B_2 \rangle + \langle A_3 B_2 \rangle\nonumber\\
& + \langle A_3 B_3 \rangle - \langle A_1 B_3 \rangle| \leq 4.
\label{chain3}
\end{align}
On the other hand, Chained $4$-settings Bell-CHSH inequality \cite{chain1, chain2, chain3} has the following form,
\begin{align}
Chain^4 &= |\langle A_1 B_1 \rangle + \langle A_2 B_1 \rangle + \langle A_2 B_2 \rangle + \langle A_3 B_2 \rangle + \langle A_3 B_3 \rangle \nonumber\\
&+ \langle A_4 B_3 \rangle + \langle A_4 B_4 \rangle - \langle A_1 B_4 \rangle| \leq 6.
\label{chain4}
\end{align}
Interestingly, using chained Bell-CHSH inequality it was shown that nonlocality and entanglement are not only different properties but are inversely related in certain scenarios\cite{chap3}. Chained Bell-CHSH inequality was shown to be increasingly sensitive to any noise in the system as the number of measurement settings increases in the inequality \cite{chap1}. It also allows to put strong lower bounds on the nonlocal content of an observed statistics \cite{chap1}. This inequality was also used for self-testing protocols \cite{chap4}. Chained Bell-CHSH inequality was shown to be necessary for probing nonlocality in experiments using energy-time and time-bin entanglement \cite{chap2}.

Another set of Bell-CHSH inequalities for $n$ dichotomic measurements performed by each party was derived by Gisin in Ref. \cite{gisin}. Bell-CHSH inequality for $3$-settings derived by Gisin \cite{gisin} is given by,
\begin{align}
G^3 & = |\langle A_1 B_1 \rangle + \langle A_1 B_2 \rangle + \langle A_1 B_3 \rangle  + \langle A_2 B_1 \rangle + \langle A_2 B_2 \rangle \nonumber\\
& - \langle A_2 B_3 \rangle + \langle A_3 B_1 \rangle - \langle A_3 B_2 \rangle - \langle A_3 B_3 \rangle| \leq 5.
\label{gisin3}
\end{align}
We will call this inequality as $3$-settings Gisin inequality. Similarly, the Bell-CHSH inequality for $4$-settings derived by Gisin \cite{gisin} is given by,
\begin{align}
G^4 & = |\langle A_1 B_1 \rangle + \langle A_1 B_2 \rangle + \langle A_1 B_3 \rangle  + \langle A_1 B_4 \rangle + \langle A_2 B_1 \rangle \nonumber\\
& + \langle A_2 B_2 \rangle + \langle A_2 B_3 \rangle - \langle A_2 B_4 \rangle + \langle A_3 B_1 \rangle + \langle A_3 B_2 \rangle \nonumber\\
& -  \langle A_3 B_3 \rangle -  \langle A_3 B_4 \rangle + \langle A_4 B_1 \rangle - \langle A_4 B_2 \rangle - \langle A_4 B_3 \rangle \nonumber \\
& - \langle A_4 B_4 \rangle | \leq 8.
\label{gisin4}
\end{align}
We will call this inequality as $4$-settings Gisin inequality. Using these inequalities it has been demonstrated that the ratios of their maximum quantum violations and their local realist bounds tend to a constant value in the limit of arbitrarily large number of measurement settings per party \cite{gisin}.

Another form of local realist inequality was derived by Collins et. al., involving three dichotomic measurements per party \cite{i3322}. This inequality is known as $I_{3322}$ inequality. $I_{3322}$ inequality has the following form,
\begin{align}
I_{3322} &= P (A_1 = +, B_1 = +) + P (A_2 = +, B_1 = +) \nonumber\\
&+ P (A_3 = +, B_1 = +) + P (A_1 = +, B_2 = +)  \nonumber\\
&+ P (A_2 = +, B_2 = +) - P (A_3 = +, B_2 = +)  \nonumber\\
&+ P (A_1 = +, B_3 = +) - P (A_2 = +, B_3 = +)  \nonumber\\
&- P (A_1 = +) - 2 P (B_1 = +) - P (B_2 = +) \leq 0 .
\label{i3322}
\end{align}
It has been shown that there exists states that violate the above inequality, but do not violate the CHSH inequality. In other words, this inequality is `relevant' to CHSH inequality \cite{i3322}. Using this inequality also, it has been demonstrated that nonlocality and entanglement are inversely related \cite{chap3}.

Deng et. al. derived another local realist inequality involving four dichotomic measurements per party \cite{deng}. This inequality is `relevant' to CHSH inequality and $I_{3322}$ inequality in the sense that there exists states that violate the inequality by Deng et. al., but do not violate the CHSH inequality or $I_{3322}$ inequality and vice-versa. $4$ settings inequality derived by Deng et. al. \cite{deng} has the form given by,
\begin{align}
D^4 & = |\langle A_1 B_1 \rangle + \langle A_2 B_2 \rangle + \langle A_1 B_2 \rangle  + \langle A_2 B_1 \rangle + \langle A_1 B_4 \rangle \nonumber\\
& + \langle A_4 B_1 \rangle - \langle A_2 B_4 \rangle - \langle A_4 B_2 \rangle - 2 \langle A_3 B_3 \rangle + \langle A_3 B_1 \rangle \nonumber\\
& + \langle A_1 B_3 \rangle +  \langle A_3 B_2 \rangle + \langle A_2 B_3 \rangle  | \leq 6.
\label{deng4}
\end{align}
We will call this inequality as DZC (Deng-Zhou-Chen) inequality.

Brunner et. al. derived a series of local realist inequalities in the scenario where each party performs four dichotomic measurements \cite{bru}. $4$ settings inequality derived by Brunner et. al., which will be used by us in the present study, has the form given by \cite{bru},
\begin{align}
B^4 &= P (A_1 = +, B_1 = +) + P (A_2 = +, B_1 = +) \nonumber\\
&+ P (A_3 = +, B_1 = +) + P (A_4 = +, B_1 = +)  \nonumber\\
&+ P (A_1 = +, B_2 = +) + P (A_2 = +, B_2 = +)  \nonumber\\
&+ P (A_3 = +, B_2 = +) - P (A_4 = +, B_2 = +)  \nonumber\\
&+ P (A_1 = +, B_3 = +) + P (A_2 = +, B_3 = +)  \nonumber\\
&- 2 P (A_3 = +, B_3 = +) + P (A_1 = +, B_4 = +)  \nonumber\\
& - P (A_2 = +, B_4 = +) - 2 P (A_1 = +)  -  P (A_2 = +) \nonumber\\
& - 2 P (B_1 = +) - P (B_2 = +) \leq 0 .
\label{bru4}
\end{align}
We will call this inequality as BG (Brunner-Gisin) inequality.

Besides these, two more inequivalent local realist inequalities in the scenario where each party performs four dichotomic measurements were derived by Avis et. al. \cite{avis} and Gisin \cite{gisin_open}, which are shown to be inequivalent to CHSH inequality \cite{avis}. One of these inequalities has the form given by \cite{avis, gisin_open},
\begin{align}
A_1^4 &= |2 \langle A_1 B_1 \rangle + \langle A_1 B_2 \rangle + \langle A_1 B_3 \rangle  + 2 \langle A_1 B_4 \rangle + \langle A_2 B_1 \rangle \nonumber\\
& + \langle A_2 B_2 \rangle + 2 \langle A_2 B_3 \rangle - 2 \langle A_2 B_4 \rangle + \langle A_3 B_1 \rangle + 2 \langle A_3 B_2 \rangle \nonumber\\
& - 2 \langle A_3 B_3 \rangle -  \langle A_3 B_4 \rangle + 2 \langle A_4 B_1 \rangle - 2  \langle A_4 B_2 \rangle -  \langle A_4 B_3 \rangle \nonumber\\
& -  \langle A_4 B_4 \rangle | \leq 10.
\label{avis1}
\end{align}
We will call this inequality as 1st AIIG (Avis-Imai-Ito-Gisin) inequality. Another inequality derived by Avis et. al. and Gisin has the form given by \cite{avis, gisin_open},
\begin{align}
A_2^4 &= |2 \langle A_1 B_1 \rangle + \langle A_1 B_2 \rangle  +  \langle A_1 B_4 \rangle 
+ \langle A_2 B_1 \rangle - \langle A_2 B_2 \rangle \nonumber\\
& +  \langle A_2 B_3 \rangle -  \langle A_2 B_4 \rangle 
 +  \langle A_3 B_2 \rangle  -  \langle A_3 B_4 \rangle 
 +  \langle A_4 B_1 \rangle \nonumber\\
& -   \langle A_4 B_2 \rangle -  \langle A_4 B_3 \rangle  -  \langle A_4 B_4 \rangle | \leq 6.
\label{avis2}
\end{align}
We will call this inequality as 2nd AIIG (Avis-Imai-Ito-Gisin) inequality.

We will use these local realist inequalities later for the purpose of the present study.

\section{Setting up the measurement scenario}\label{sec3}

In order to probe sharing of bipartite nonlocality we consider the following measurement scenario in the present study:

\begin{scenario} \label{scenario1}
Two spin-$\frac{1}{2}$ particles are prepared in a bipartite state $\rho$ and they are spatially separated. These two particles are shared by Alice and multiple Bobs (say, Bob$^1$, Bob$^2$, Bob$^3$, ..., Bob$^n$), where Alice performs measurements on the first particle and multiple Bobs perform measurements on the second particle sequentially. After doing measurements on the second particle Bob$^1$ delivers the particle to Bob$^2$, and similarly, Bob$^2$ passes the second particle to Bob$^3$ after  measurements, and so on.
\end{scenario}

In Scenario \ref{scenario1} we consider the following two salient features:

1) Each Bob performs measurements independent of the measurement settings and outcomes of the previous Bobs on the particle in his possession.

2) all possible measurement settings of each Bob are equally probable, i. e., we are considering unbiased input scenario for each Bob.

\begin{figure}
	\centering
	\includegraphics[width=0.45\textwidth]{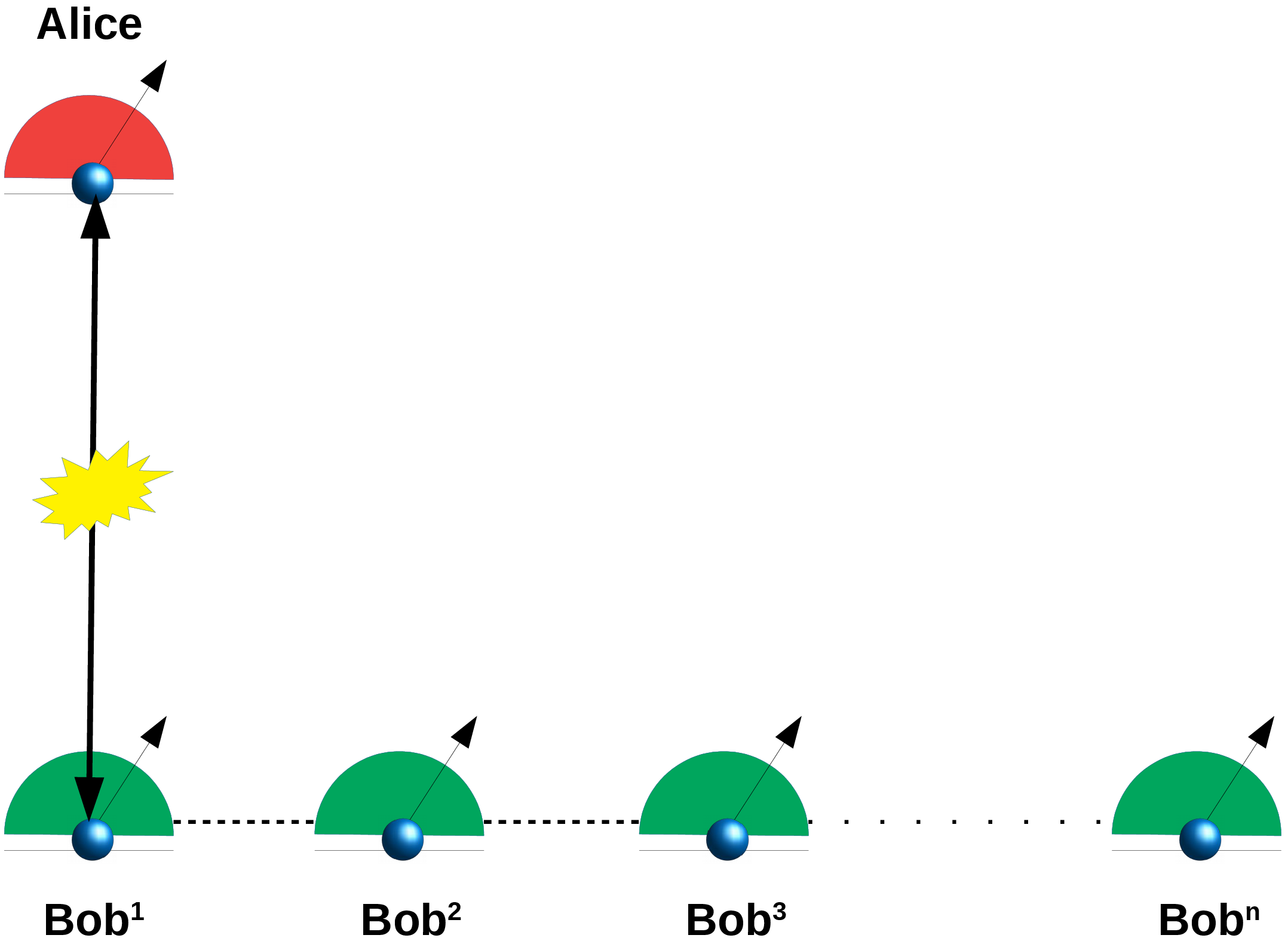}
	\caption{Solid black line indicates spatial separation and dashed black line indicates time-separation. Alice performs measurements on one particle and multiple Bobs (Bob$^1$, Bob$^2$, Bob$^3$, ..., Bob$^n$) perform measurements on the other particle sequentially.}\label{fig1}
\end{figure}

Note that in this Scenario \ref{scenario1} no-signaling condition is satisfied between Alice and any Bob as they are spatially separated and perform measurements on two different particles. On the other hand, in this scenario no-signaling condition is not satisfied between different Bobs as Bob$^1$ implicitly signals to Bob$^2$ by his choice of measurement on the state before he passes it on. Similarly, Bob$^2$ also signals to Bob$^3$ and so on. This Scenario \ref{scenario1} is depicted in Fig. \ref{fig1}.

In Scenario \ref{scenario1} we investigate how many Bobs can sequentially demonstrate nonlocality with single Alice with respect to the quantum violations of different local realist inequalities.  As we want to explore how many Bobs can have measurement statistics violating different local realist inequalities with a single Alice, each of the Bobs except the last Bob cannot perform projective measurements. If any Bob performs projective measurements, there would be no possibility of violation of the local realist inequalities by the next Bob, since  the entanglement of the state shared between Alice and 
the subsequent Bob would be completely destroyed. Hence, in order to address the aforementioned question with $n$ Bobs, the measurements of the first $(n -1)$ Bobs should be weak.

\textbf{Description of fuzzy measurement}: 

Weak measurement without post selection or fuzzy is characterised by two parameters: 1) the quality factor $F$ which quantifies the extent to which the state of the system remains undisturbed after the measurement, and 2) the precision $G$ of the measurements which quantifies the information gain from measurement \cite{sygp}. In case of strong projective measurement, $F = 0$ and $G = 1$. An optimal pointer state is defined as the one which gives the best trade-off between these two quantities, i.e., for a given quality factor, it provides the greatest precision. It has been shown that the information-disturbance trade-off condition for an optimal pointer is given by, $F^2 + G^2 =1$ \cite{sygp}. 

This weak measurement formalism has been recast in the unsharp measurement formalism in Ref. \cite{mal, sas}, which is a particular class of positive operator valued measurement (POVM) \cite{pb1, pb2}. POVM is a set of positive operators that add up to identity, i. e.,  $E \equiv \{ E_i | \sum_i E_i = \mathbb{I}, 0 < E_i \leq \mathbb{I} \}$. Effects represent quantum events that may occur as outcomes of a measurement. In case of a dichotomic unsharp measurement, the effect operators are given by, $E^\lambda_{\pm}$ = $(\mathbb{I}_2 \hspace{0.1cm} \pm \hspace{0.1cm} \lambda \hat{n} \cdot \vec{\sigma})/2$. Here $\vec{\sigma}$ = $(\sigma_1, \sigma_2, \sigma_3)$ is a vector composed of Pauli matrices, $\hat{n}$ is a three dimensional unit vector, $\mathbb{I}_2$ is the $2 \times 2$ identity matrix, $\lambda$ is the sharpness parameter, $0 < \lambda \leq 1$. The probability of getting the outcomes $+$ and $-$ are $\tr[E^\lambda_{+} \rho]$ and $\tr[E^\lambda_{-} \rho]$, respectively, where $\rho$ is the state of the system on which unsharp measurement is performed. The post-measurement states, when the outcomes $+$ and $-$ are obtained, are  determined by Von Neumann-Luder's transformation rule \cite{pb1}, upto local unitary freedom, as $\dfrac{\sqrt{E^\lambda_{+}} \rho \sqrt{E^\lambda_{+}}}{\tr[E^\lambda_{+} \rho]}$ and $\dfrac{\sqrt{E^\lambda_{-}} \rho \sqrt{E^\lambda_{-}}}{\tr[E^\lambda_{-} \rho]}$, respectively.

It was shown in \cite{mal} that weak measurement formalism characterised by $F$ and $G$ is related to the unsharp measurement formalism through the relations given by, $F = \sqrt{1-\lambda^2}$ and $G = \lambda$. Hence, $\lambda$ characterizes the precision of the measurement. For strong projective measurement, $G = \lambda = 1$, $F=0$. It can be easily checked that the optimal pointer state condition, $F^2 + G^2 = 1$, is automatically satisfied in the unsharp measurement formalism \cite{mal, sas}. 

In the next Section we consider Scenario \ref{scenario1}, where multiple observers perform sequential measurements on one part of a shared bipartite entangled state.

\section{Sharing of bipartite nonlocality when each party performs three dichotomic measurements} \label{sec4}

In this Section we investigate how many Bobs can sequentially demonstrate nonlocality with single Alice in Scenario \ref{scenario1} using the chained $3$-settings Bell-CHSH inequality (\ref{chain3}), $3$-settings Gisin inequality (\ref{gisin3}) and $I_{3322}$ inequality (\ref{i3322}). In all these three inequalities each party performs three dichotomic measurements. Note that all these inequalities are maximally violated by maximally entangled state. Hence, in order to probe optimal sharing of bipartite nonlocality by multiple Bobs using the above inequalities, we consider that Alice and multiple Bobs (say, Bob$^1$, Bob$^2$, Bob$^3$, ..., Bob$^n$) initially share a maximally entangled state, say, singlet state given by,
\begin{equation}
| \psi \rangle = \frac{1}{\sqrt{2}} (|01\rangle - |10 \rangle),
\label{singlet}
\end{equation}
where $|0\rangle$ and $|1 \rangle$ form an orthonormal basis in $\mathbb{C}^2$. 
%Suppose, Alice has a choice between three dichotomic measurements: spin component observables in the directions $\{ \hat{x}^0, \hat{x}^1, \hat{x}^2 \}$,  and Bob$^i$ ($ i \in \{1, 2, 3, ..., n\}$) also has the choice between the spin component observables in the directions $\{ \hat{y}^0_i, \hat{y}^1_i, \hat{y}^2_i \}$.

We assume that the three possible choices of measurement settings of Alice are the spin component observables in the directions $\hat{x}^u$, i. e., observables corresponding to the operators $\vec{\sigma} \cdot \hat{x}^u$. Here $\vec{\sigma}$ = $(\sigma_1, \sigma_2, \sigma_3)$ is a vector composed of Pauli matrices and $\hat{x}^u$ is given by,
\begin{equation}
\label{alicedir}
\hat{x}^u = \sin \theta^x_u \cos \phi^x_u \hat{X} + \sin \theta^x_u \sin \phi^x_u \hat{Y} + \cos \theta^x_u \hat{Z},
\end{equation}
where $0 \leq \theta^x_u \leq \pi$; $0 \leq \phi^x_u \leq 2 \pi$; $u \in \{0, 1, 2\}$; $\hat{X}$, $\hat{Y}$, $\hat{Z}$ are three mutually orthogonal unit vectors in Cartesian coordinate system. The three possible choices of measurement settings of Bob$^i$ are the spin component observables in the directions $\hat{y}_i^v$, i. e., observables corresponding to the operators $\vec{\sigma} \cdot \hat{y}_i^v$. Here $\hat{y}_i^v$ is given by,
\begin{equation}
\label{bobndir}
\hat{y}_i^v = \sin \theta^{y_i}_v \cos \phi^{y_i}_v \hat{X} + \sin \theta^{y_i}_v \sin \phi^{y_i}_v \hat{Y} + \cos \theta^{y_i}_v \hat{Z},
\end{equation}
where $0 \leq \theta^{y_i}_v \leq \pi$; $0 \leq \phi^{y_i}_v \leq 2 \pi$; $v \in \{0, 1, 2\}$ and  $i\in\{1, 2, 3, ..., n\}$. Outcomes of these measurements are labelled by $\{-1, +1 \}$.

The correlation function between Alice and Bob$^1$, when Alice performs a projective measurement of the spin component along the direction $\hat{x}^u$, and Bob$^1$ performs an unsharp measurement of the spin component along the direction $\hat{y}_1^v$, is given by,
\begin{equation}
C_1^{uv} = -\lambda_1 (\hat{y}_1^v \cdot \hat{x}^u),
\end{equation}
where $u \in \{0, 1, 2\}$; $v \in \{0, 1, 2\}$; $\lambda_1$ is the sharpness parameter of Bob$^1$'s unsharp measurements. For two Bobs measuring sequentially on the same particle, the joint probability of obtaining the outcomes $a$ and $b_2$ (where $a \in \{-1, +1 \}$ and $b_2 \in \{-1, +1 \}$) when Alice performs a projective measurement of the spin component along the direction $\hat{x}^u$ and Bob$^2$ performs an unsharp measurement of the spin component along the direction $\hat{y}_2^v$, respectively, given that Bob$^1$ has performed an unsharp measurement of the spin component along the direction $\hat{y}_1^k$, is given by,
\begin{align}
p(a, b_2|\hat{x}^u, \hat{y}_1^k, \hat{y}_2^v) &= \Big(\sqrt{1 - \lambda_1^2} \Big) \frac{ 1 - a b_2 \lambda_2 (\hat{y}_2^v \cdot \hat{x}^u)}{4}  \nonumber\\
&+ \Big(1- \sqrt{1 - \lambda_1^2}\Big) \frac{1 - a b_2 \lambda_2 (\hat{y}_1^k \cdot \hat{x}^u)(\hat{y}_2^v \cdot \hat{y}_1^k)}{4},
\end{align} 
where $\lambda_1$ and $\lambda_2$ are the sharpness parameters of the measurements performed by Bob$^1$ and Bob$^2$, respectively; $u \in \{0, 1, 2\}$; $k \in \{0, 1, 2\}$; $v \in \{0, 1, 2\}$. In this case, the correlation function between Alice and Bob$^2$ is given by,
\begin{align}
C_2^{uv} =& -\lambda_2 \Big[ \Big(\sqrt{1-\lambda_1^2}\Big) (\hat{y}_2^v \cdot \hat{x}^u) \nonumber\\
&+ \Big(1- \sqrt{1-\lambda_1^2} \Big)  (\hat{y}_1^k \cdot \hat{x}^u)  (\hat{y}_2^v \cdot \hat{y}_1^k)\Big].
\label{corr}
\end{align}
Since we have assumed that Bob$^2$ is ignorant about the measurement settings of Bob$^1$, the correlation (\ref{corr}) has to be averaged over the three possible choices of measurement settings performed by Bob$^1$ (i. e., the spin component observables in the directions $\{ \hat{y}^0_1, \hat{y}^1_1, \hat{y}^2_1 \}$). This average correlation function between Alice and Bob$^2$ is given by,
\begin{equation}
\overline{C_2^{uv}} = \sum_{k = 0,1,2} C_2^{uv} P(\hat{y}_1^k),
\label{avcorr}
\end{equation}
where $P(\hat{y}_1^k)$ is the probability with which Bob$^1$ measures spin component observables in the direction $\hat{y}_1^k$. Since we are considering an unbiased input scenario, all the possible measurement settings of the previous Bob are equally probable, i. e.,  $P(\hat{y}_1^0)$ = $P(\hat{y}_1^1)$ = $P(\hat{y}_1^2)$ = $\frac{1}{3}$. Hence, from Eqs. (\ref{corr}, \ref{avcorr}) we obtain,
\begin{align}
\overline{C_2^{uv}} =& -\frac{\lambda_2}{3}\Big[\sqrt{1-\lambda_1^2} (\hat{y}_2^v \cdot \hat{x}^u) + \Big(1- \sqrt{1-\lambda_1^2}\Big)  (\hat{y}_1^0 \cdot \hat{x}^u)  (\hat{y}_2^v \cdot \hat{y}_1^0)\Big] \nonumber\\
& -\frac{\lambda_2}{3}\Big[\sqrt{1-\lambda_1^2} (\hat{y}_2^v \cdot \hat{x}^u) + \Big(1- \sqrt{1-\lambda_1^2}\Big)  (\hat{y}_1^1 \cdot \hat{x}^u)  (\hat{y}_2^v \cdot \hat{y}_1^1)\Big] \nonumber\\
&-\frac{\lambda_2}{3}\Big[\sqrt{1-\lambda_1^2} (\hat{y}_2^v \cdot \hat{x}^u) + \Big(1- \sqrt{1-\lambda_1^2}\Big)  (\hat{y}_1^2 \cdot \hat{x}^u)  (\hat{y}_2^v \cdot \hat{y}_1^2)\Big].
\end{align}
In a similar way, the average correlation functions between Alice and Bob$^i$, $\overline{C_i^{uv}}$ (where $u \in \{0, 1, 2\}$, $v \in \{0, 1, 2\}$), can be evaluated.

Using such average correlations between Alice and Bob$^i$, we get the following form of chained $3$-settings Bell-CHSH inequality (\ref{chain3}) for Alice and Bob$^i$.
\begin{equation}
Chain^3_i = |\overline{C_i^{00}}  + \overline{C_i^{10}}  + \overline{C_i^{11}} + \overline{C_i^{21}} + \overline{C_i^{22}} - \overline{C_i^{02}} | \leq 4.
\label{chain3i}
\end{equation}

Now consider whether Bob$^1$ and Bob$^2$ can sequentially demonstrate bipartite nonlocality with single Alice through the quantum violations of chained $3$-settings Bell-CHSH inequality (\ref{chain3i}). The measurements of the final Bob (i. e., Bob$^2$) are sharp ($\lambda_2 = 1$) and the measurements of Bob$^1$ are unsharp. We observe that when Bob$^1$ gets $5\%$ violation of the chained $3$-settings Bell-CHSH inequality (\ref{chain3i}), i. e., when $Chain^3_1 = 4.20$, then the maximum quantum mechanical violation of chained $3$-settings Bell-CHSH inequality (\ref{chain3i}) for Bob$^2$ is $3. 25 \%$, i.e., $Chain^3_2 = 4.13$. This happens for the choice of measurement settings: ($\theta^x_0$, $\phi^x_0$, $\theta^x_1$, $\phi^x_1$, $\theta^x_2$, $\phi^x_2$, $\theta^{y_1}_0$, $\phi^{y_1}_0$, $\theta^{y_1}_1$, $\phi^{y_1}_1$, $\theta^{y_1}_2$, $\phi^{y_1}_2$, $\theta^{y_2}_0$, $\phi^{y_2}_0$, $\theta^{y_2}_1$, $\phi^{y_2}_1$, $\theta^{y_2}_2$, $\phi^{y_2}_2$) $\equiv$ ($1.19$, $1.13$, $1.21$, $6.28$, $1.59$, $5.28$, $2.00$, $3.70$, $1.76$, $2.62$, $1.34$, $1.66$, $2.00$, $3.70$, $1.76$, $2.62$, $1.34$, $1.66$) with $\lambda_1 =0.81$. In fact, it is observed that for $\lambda_1 \in[0.77, 0.84]$ both Bob$^1$ and Bob$^2$ can sequentially demonstrate bipartite nonlocality through the quantum violations of chained $3$-settings Bell-CHSH inequality (\ref{chain3i}).

Next, we  address the question whether Bob$^1$, Bob$^2$ and Bob$^3$ can sequentially demonstrate bipartite nonlocality with single Alice through the quantum violations of chained $3$-settings Bell-CHSH inequality (\ref{chain3i}). In this case, the measurements of the final Bob (i. e., Bob$^3$) are sharp ($\lambda_3 = 1$), and the measurements of Bob$^1$ and Bob$^2$ are unsharp. Here we observe that, when each of the quantum violation of chained $3$-settings Bell-CHSH inequality (\ref{chain3i}) for Bob$^1$ and Alice and that for Bob$^2$ and Alice is $2.5\%$, i.e., when $Chain^3_1 = 4.10$ and $Chain^3_2 = 4.10$, then the maximum quantum value of the left hand side of chained $3$-settings Bell-CHSH inequality (\ref{chain3i}) for Bob$^3$ and Alice is given by, $Chain^3_3 = 2.54$. In fact, when $Chain^3_1 = 4$, $Chain^3_2 = 4$, then the maximum quantum value of $Chain^3_3 = 2.86$. This happens for the choice of measurement settings: ($\theta^x_0$, $\phi^x_0$, $\theta^x_1$, $\phi^x_1$, $\theta^x_2$, $\phi^x_2$, $\theta^{y_1}_0$, $\phi^{y_1}_0$, $\theta^{y_1}_1$, $\phi^{y_1}_1$, $\theta^{y_1}_2$, $\phi^{y_1}_2$, $\theta^{y_2}_0$, $\phi^{y_2}_0$, $\theta^{y_2}_1$, $\phi^{y_2}_1$, $\theta^{y_2}_2$, $\phi^{y_2}_2$, $\theta^{y_3}_0$, $\phi^{y_3}_0$, $\theta^{y_3}_1$, $\phi^{y_3}_1$, $\theta^{y_3}_2$, $\phi^{y_3}_2$) $\equiv$ ($1.19$, $1.13$, $1.21$, $6.28$, $1.59$, $5.28$, $2.00$, $3.70$, $1.76$, $2.62$, $1.34$, $1.66$, $2.00$, $3.70$, $1.76$, $2.62$, $1.34$, $1.66$, $2.00$, $3.70$, $1.76$, $2.62$, $1.34$, $1.66$) with $\lambda_1=0.77$ and $\lambda_2=0.94$. Hence, Bob$^1$, Bob$^2$ and Bob$^3$ cannot sequentially demonstrate bipartite nonlocality with single Alice through the quantum violations of chained $3$-settings Bell-CHSH inequality (\ref{chain3i}).

It is to be noted here that Bob$^3$ may obtain quantum mechanical violation of the chained $3$-settings Bell-CHSH inequality (\ref{chain3i}) if the sharpness parameter of Bob$^2$ is too small to get a violation. In fact, it can be easily checked that at most two Bobs (not necessarily Bob$^1$ and Bob$^2$, but any two Bobs) can sequentially demonstrate bipartite nonlocality through the quantum violations of chained $3$-settings Bell-CHSH inequality (\ref{chain3i}).

Proceeding in a similar way we have also calculated how many Bobs can sequentially demonstrate bipartite nonlocality with single Alice in Scenario \ref{scenario1} through the quantum violations of $3$-settings Gisin inequality (\ref{gisin3}) and $I_{3322}$ inequality (\ref{i3322}). One important point to be stressed here is that the left hand side of $I_{3322}$ inequality (\ref{i3322}) is a linear combination of joint probabilities and marginal probabilities. Hence, in order to evaluate the quantum violation of $I_{3322}$ inequality (\ref{i3322}) with Alice and Bob$^i$, the joint probabilities and the marginal probabilities at Bob$^i$'s end have to be averaged over all possible equally probable measurement settings of previous Bobs. The results obtained are summarized in Table \ref{tab1}. 

\begin{table}
	\centering
	\begin{tabular}{ |c|c| } 
		\hline
		\textit{\textbf{Local realist inequalities}} & \textit{\textbf{Maximum number of Bobs}} \\
		\textit{\textbf{}} & \textit{\textbf{who can simultaneously}} \\
		\textit{\textbf{}} & \textit{\textbf{demonstrate bipartite}} \\
		\textit{\textbf{}} & \textit{\textbf{nonlocality with single}} \\
		\textit{\textbf{}} & \textit{\textbf{Alice}} \\
		\hline
		\hline
		Chained $3$-settings & $2$  \\
		Bell-CHSH inequality (\ref{chain3}) &   \\
		\hline
		$3$-settings Gisin & $1$  \\
		inequality (\ref{gisin3}) &   \\
		\hline
		$I_{3322}$ inequality (\ref{i3322}) & $1$  \\
		\hline
	\end{tabular}
	\caption{Maximum number of Bobs who can simultaneously demonstrate bipartite nonlocality with single Alice through the quantum violations of chained $3$-settings Bell-CHSH inequality (\ref{chain3}), $3$-settings Gisin inequality (\ref{gisin3}) and $I_{3322}$ inequality (\ref{i3322}).}
	\label{tab1}
\end{table}

Table \ref{tab1} clearly indicates that the maximum number of Bobs who can sequentially demonstrate bipartite nonlocality with single Alice does not increase when each party performs three dichotomic measurements, instead of two dichotomic measurements. Now we are interested to study the above issue when each party performs four dichotomic measurements.

\section{Sharing of bipartite nonlocality when each party performs four dichotomic measurements} \label{sec5}

We now investigate how many Bobs can sequentially demonstrate nonlocality with single Alice in Scenario \ref{scenario1} using the chained $4$-settings Bell-CHSH inequality (\ref{chain4}), $4$-settings Gisin inequality (\ref{gisin4}), DZC  inequality (\ref{deng4}), BG inequality (\ref{bru4}), 1st AIIG inequality (\ref{avis1}), 2nd AIIG inequality (\ref{avis2}). In all these inequalities each party performs four dichotomic measurements. Since all these inequalities are maximally violated by maximally entangled state, we assume that Alice and multiple Bobs (say, Bob$^1$, Bob$^2$, Bob$^3$, ..., Bob$^n$) initially share a maximally entangled state, say, singlet state (\ref{singlet}). Suppose, Alice has a choice between four dichotomic measurements: spin component observables in the directions $\hat{x}^u$ (where $u \in \{0, 1, 2, 3\}$),  and Bob$^i$ ($ i \in \{1, 2, 3, ..., n\}$)  has the choice between the spin component observables in the directions $\hat{y}^v_i$ (where $v \in \{ 0, 1, 2, 3\}$) to measure. Outcomes of these measurements are labelled by $\{-1, +1 \}$. $\hat{x}^u$ and $\hat{y}^v_i$ have the form mentioned in Eq.(\ref{alicedir}) and in Eq.(\ref{bobndir}), respectively.

In this case also we assume unbiased input scenario, i. e., all three possible  measurement settings of the previous Bobs are equally probable with probability $\frac{1}{4}$. Hence, the average correlation function between Alice and Bob$^2$, when Alice performs a projective measurement of the spin component along the direction $\hat{x}^u$, and Bob$^2$ performs an unsharp measurement of the spin component along the direction $\hat{y}_2^v$, is given by,
\begin{align}
\overline{C_2^{uv}} =& -\frac{\lambda_2}{4}\Big[\sqrt{1-\lambda_1^2} (\hat{y}_2^v \cdot \hat{x}^u) + \Big(1- \sqrt{1-\lambda_1^2}\Big)  (\hat{y}_1^0 \cdot \hat{x}^u)  (\hat{y}_2^v \cdot \hat{y}_1^0)\Big] \nonumber\\
& -\frac{\lambda_2}{4}\Big[\sqrt{1-\lambda_1^2} (\hat{y}_2^v \cdot \hat{x}^u) + \Big(1- \sqrt{1-\lambda_1^2}\Big)  (\hat{y}_1^1 \cdot \hat{x}^u)  (\hat{y}_2^v \cdot \hat{y}_1^1)\Big] \nonumber\\
&-\frac{\lambda_2}{4}\Big[\sqrt{1-\lambda_1^2} (\hat{y}_2^v \cdot \hat{x}^u) + \Big(1- \sqrt{1-\lambda_1^2}\Big)  (\hat{y}_1^2 \cdot \hat{x}^u)  (\hat{y}_2^v \cdot \hat{y}_1^2)\Big]\nonumber\\
&-\frac{\lambda_2}{4}\Big[\sqrt{1-\lambda_1^2} (\hat{y}_2^v \cdot \hat{x}^u) + \Big(1- \sqrt{1-\lambda_1^2}\Big)  (\hat{y}_1^3 \cdot \hat{x}^u)  (\hat{y}_2^v \cdot \hat{y}_1^3)\Big].
\end{align}
In a similar way, the average correlation functions between Alice and  Bob$^i$, $\overline{C_i^{uv}}$ (where $u \in \{0, 1, 2, 3\}$, $v \in \{0, 1, 2, 3\}$), can be evaluated.

Using such average correlations between Alice and Bob$^i$, we get the following form of chained $4$-settings Bell-CHSH inequality (\ref{chain4}) for Alice and Bob$^i$.
\begin{equation}
Chain^4_i = |\overline{C_i^{00}}  + \overline{C_i^{10}}  + \overline{C_i^{11}} + \overline{C_i^{21}} + \overline{C_i^{22}} + \overline{C_i^{32}} + \overline{C_i^{33}} - \overline{C_i^{03}} | \leq 6.
\label{chain4i}
\end{equation}

Now consider whether both Bob$^1$ and Bob$^2$ can demonstrate bipartite nonlocality with single Alice through the quantum violations of chained $4$-settings Bell-CHSH inequality (\ref{chain4i}). In this case, the measurements of the final Bob (i. e., Bob$^2$) are sharp ($\lambda_2 = 1$) and the measurements of Bob$^1$ are unsharp. We observe that when Bob$^1$ gets $5\%$ violation of the chained $4$-settings Bell-CHSH inequality (\ref{chain4i}), i. e., when $Chain^4_1 = 6.30$, then the maximum quantum value of chained $4$-settings Bell-CHSH inequality (\ref{chain4i}) for Alice and Bob$^2$ is $Chain^4_2 = 5.63$. In fact, when $Chain^4_1 = 6$, then the maximum quantum value of chained $4$-settings Bell-CHSH inequality (\ref{chain4i}) for Alice and Bob$^2$ is $Chain^4_2 = 5.85$. This happens for the choice of measurement settings: ($\theta^x_0$, $\phi^x_0$, $\theta^x_1$, $\phi^x_1$, $\theta^x_2$, $\phi^x_2$, $\theta^x_3$, $\phi^x_3$, $\theta^{y_1}_0$, $\phi^{y_1}_0$, $\theta^{y_1}_1$, $\phi^{y_1}_1$, $\theta^{y_1}_2$, $\phi^{y_1}_2$, $\theta^{y_1}_3$, $\phi^{y_1}_3$, $\theta^{y_2}_0$, $\phi^{y_2}_0$, $\theta^{y_2}_1$, $\phi^{y_2}_1$, $\theta^{y_2}_2$, $\phi^{y_2}_2$, $\theta^{y_2}_3$, $\phi^{y_2}_3$) $\equiv$ ($0.39$, $0$, $1.18$, $0$, $1.96$, $6.28$, $2.75$, $0$, $2.36$, $3.14$, $1.57$, $3.14$, $0.78$, $3.14$, $0$, $5.77$, $2.36$, $3.14$, $1.57$, $3.14$, $0.78$, $3.14$, $0$, $6.09$) with $\lambda_1 =0.81$. Hence, both Bob$^1$ and  Bob$^2$ cannot demonstrate bipartite nonlocality with single Alice through the quantum violations of chained $4$-settings Bell-CHSH inequality (\ref{chain4i}).

Note that Bob$^2$ may get quantum violation of the chained $4$-settings Bell-CHSH inequality (\ref{chain4i}) if the sharpness parameter $\lambda_1$ of Bob$^1$ is too small such that Bob$^1$ does not get any violation. In fact, at most one Bob (not necessarily Bob$^1$, but any one Bob) can show bipartite nonlocality with single Alice through the quantum violation of chained $4$-settings Bell-CHSH inequality (\ref{chain4i}).

Following similar approach we have investigated how many Bobs can sequentially demonstrate bipartite nonlocality with single Alice in Scenario \ref{scenario1} through the quantum violations of $4$-settings Gisin inequality (\ref{gisin4}), DZC  inequality (\ref{deng4}), BG inequality (\ref{bru4}), 1st AIIG inequality (\ref{avis1}), 2nd AIIG inequality (\ref{avis2}). The results obtained are summarized in Table \ref{tab2}.

\begin{table}
	\centering
	\begin{tabular}{ |c|c| } 
		\hline
		\textit{\textbf{Local realist inequalities}} & \textit{\textbf{Maximum number of Bobs}} \\
		\textit{\textbf{}} & \textit{\textbf{who can simultaneously}} \\
		\textit{\textbf{}} & \textit{\textbf{demonstrate bipartite}} \\
		\textit{\textbf{}} & \textit{\textbf{nonlocality with single}} \\
		\textit{\textbf{}} & \textit{\textbf{Alice}} \\
		\hline
		\hline
		Chained $4$-settings & $1$  \\
		Bell-CHSH inequality (\ref{chain4}) &   \\
		\hline
		$4$-settings Gisin & $2$  \\
		inequality (\ref{gisin4}) &   \\
		\hline
		DZC  inequality (\ref{deng4}) & $2$  \\
		\hline
		BG inequality (\ref{bru4}) & $2$  \\
		\hline
		1st AIIG inequality (\ref{avis1}) & $2$  \\
		\hline
		2nd AIIG inequality (\ref{avis2}) & $2$  \\
		\hline
	\end{tabular}
	\caption{Maximum number of Bobs who can simultaneously demonstrate bipartite nonlocality with single Alice through the quantum violations of chained $4$-settings Bell-CHSH inequality (\ref{chain4}), $4$-settings Gisin inequality (\ref{gisin4}), DZC  inequality (\ref{deng4}), BG inequality (\ref{bru4}), 1st AIIG inequality (\ref{avis1}), 2nd AIIG inequality (\ref{avis2}).}
	\label{tab2}
\end{table}

From Table \ref{tab2} it is evident that no advantage is gained in the context of sharing of nonlocality in Scenario \ref{scenario1} when one uses local realist inequalities with four dichotomic measurements per party, instead of using CHSH inequality. From these results we conjecture the following:

\begin{conjecture}
At most two Bobs can demonstrate bipartite nonlocality with single Alice in Scenario \ref{scenario1} through the quantum violations of local realist inequalities that use $n$ dichotomic measurements per party, where $n$ is arbitrary.
\end{conjecture}

From Tables \ref{tab1}, \ref{tab2} we observe that at most two Bobs can demonstrate bipartite nonlocality with single Alice in Scenario \ref{scenario1} through the quantum violations of chained $3$-settings Bell-CHSH inequality (\ref{chain3}), $4$-settings Gisin inequality (\ref{gisin4}), DZC  inequality (\ref{deng4}), BG inequality (\ref{bru4}), 1st AIIG inequality (\ref{avis1}), 2nd AIIG inequality (\ref{avis2}), which was also probed using CHSH inequality (\ref{chsh}) \cite{sygp, mal, exp1, exp2}. Note that in all these studies it is  assumed that maximally entangled pure state is initially shared between Alice and multiple Bobs. Hence, it is legitimate ask whether the aforementioned six local realist inequalities give any advantage over CHSH inequality in the context of sharing of bipartite nonlocality in Scenario \ref{scenario1} when non-maximally entangled pure state or mixed state is initially shared between Alice and multiple Bobs.

\section{Sharing of bipartite nonlocality when the initial state is non-maximally entangled pure state or mixed state} \label{sec6}

\begin{table}
	\centering
\begin{tabular}{ |c|c| } 
 \hline
\textit{\textbf{Local realist inequalities}} & \textit{\textbf{$C_{min}$}} \\
\hline
\hline
CHSH inequality (\ref{chsh}) &  $0.76$ \\
\hline
Chained $3$-settings Bell-CHSH inequality (\ref{chain3}) & $0.92$  \\
\hline
$4$-settings Gisin  inequality (\ref{gisin4}) & $0.91$  \\
\hline
DZC  inequality (\ref{deng4}) & $0.82$  \\
\hline
BG inequality (\ref{bru4}) & $0.82$  \\
\hline
1st AIIG inequality (\ref{avis1}) & $0.84$  \\
\hline
2nd AIIG inequality (\ref{avis2}) & $0.82$  \\
\hline
\end{tabular}
\caption{$C_{min}$ for CHSH inequality (\ref{chsh}), chained $3$-settings Bell-CHSH inequality (\ref{chain3}), $4$-settings Gisin inequality (\ref{gisin4}), DZC  inequality (\ref{deng4}), BG inequality (\ref{bru4}), 1st AIIG inequality (\ref{avis1}), 2nd AIIG inequality (\ref{avis2}).}
\label{tab3}
\end{table}

In previous Sections we have considered that maximally entangled pure state (singlet state) is shared between Alice and multiple Bobs. However, in practical situation it is very difficult to prepare maximally entangled pure state. Hence, in order to consider the imperfection that may appear in real experimental scenario, we consider that Alice and multiple Bobs initially share non-maximally entangled pure state, instead of singlet state. Any pure two-qubit state can be written in the following form, called the Schmidt decomposition \cite{schmidt1, schmidt2},
\begin{equation}
|\psi (\alpha) \rangle = \cos \alpha |00\rangle + \sin \alpha |11 \rangle,
\label{nonmax}
\end{equation}
where $0 \leq \alpha \leq \frac{\pi}{2}$. Note that concurrence \cite{con}, which is a measure of entanglement, of the state $|\psi (\alpha) \rangle$ given by Eq.(\ref{nonmax}) turns out to be $C = \sin 2 \alpha$. $\alpha = \frac{\pi}{4}$ characterizes maximally entangled pure state. 

In previous Sections we observe that at most two Bobs can sequentially demonstrate bipartite nonlocality with single Alice in Scenario \ref{scenario1} using CHSH inequality (\ref{chsh}), chained $3$-settings Bell-CHSH inequality (\ref{chain3}), $4$-settings Gisin inequality (\ref{gisin4}), DZC  inequality (\ref{deng4}), BG inequality (\ref{bru4}), 1st AIIG inequality (\ref{avis1}), 2nd AIIG inequality (\ref{avis2}) when maximally entangled pure state (i. e., state with concurrence $C = 1$) is initially shared between Alice and multiple Bobs. However, if the concurrence of the shared state is decreased from $C=1$, then the above feature persists up to a certain value of concurrence. Below that value of concurrence of the shared state, two Bobs cannot sequentially demonstrate bipartite nonlocality with single Alice in Scenario \ref{scenario1}. Let us assume that $C_{min}$ denotes the minimum permissible value of Concurrence of the state, initially shared between Alice and multiple Bobs, for which two Bobs can sequentially demonstrate bipartite nonlocality with single Alice in Scenario \ref{scenario1}. If the concurrence of the shared state $C < C_{min}$, then two Bobs cannot sequentially demonstrate bipartite nonlocality with single Alice in Scenario \ref{scenario1}. We have calculated $C_{min}$ using the above seven local realist inequalities. The results are presented in Table \ref{tab3}.

From Table \ref{tab3} it is clear that CHSH inequality is the most robust against entanglement of the shared state in the context of sharing of bipartite nonlocality by multiple Bobs with single Alice in Scenario \ref{scenario1}. In other words, there is range of concurrence $C \in [0.76, 0.82)$  of the initial shared state for which two Bobs can sequentially demonstrate bipartite nonlocality with single Alice in Scenario \ref{scenario1} using CHSH inequality (\ref{chsh}), but not using the other local-realist inequalities considered.

Note that environmental effects may turn a pure state into a mixed one. In order to incorporate this issue we now consider that Alice and multiple Bobs initially share mixed state, instead of pure singlet state, given by,
\begin{equation}
\rho_w = w | \psi_{sing} \rangle \langle \psi_{sing} | + (1-w) \frac{\mathbb{I}_2 \otimes \mathbb{I}_2}{4}, 
\label{mixed}
\end{equation}
where $| \psi_{sing} \rangle$ = $\frac{1}{\sqrt{2}} (|01\rangle - |10 \rangle)$ is the singlet state. $(1-w)$ denotes the amount of white-noise that is mixed with the singlet state. $w=1$ implies pure state. Hence, $w$ characterizes mixedness of the state (\ref{mixed}). Similar to the previous case, in this case also, we have calculated the minimum permissible values of $w$, denoted by $w_{min}$, for which two Bobs can sequentially demonstrate bipartite nonlocality with single Alice in Scenario \ref{scenario1} using CHSH inequality (\ref{chsh}), chained $3$-settings Bell-CHSH inequality (\ref{chain3}), $4$-settings Gisin inequality (\ref{gisin4}), DZC  inequality (\ref{deng4}), BG inequality (\ref{bru4}), 1st AIIG inequality (\ref{avis1}), 2nd AIIG inequality (\ref{avis2}) when the mixed state $\rho_w$ (\ref{mixed}) is initially shared between Alice and multiple Bobs. If the state $\rho_w$ (\ref{mixed}) with $w \geq w_{min}$ is initially shared between Alice and multiple Bobs, then at most two Bobs can sequentially demonstrate bipartite nonlocality with single Alice in Scenario \ref{scenario1} using the respective local-realist inequality. On the other hand, if the state $\rho_w$ (\ref{mixed}) with $w < w_{min}$ is initially shared between Alice and multiple Bobs, then two Bobs cannot sequentially demonstrate bipartite nonlocality with single Alice in Scenario \ref{scenario1} using the respective local-realist inequality. We have calculated $w_{min}$ using the above seven local realist inequalities. The results are presented in Table \ref{tab4}.

\begin{table}[t]
	\centering
	\begin{tabular}{ |c|c| } 
		\hline
		\textit{\textbf{Local realist inequalities}} & \textit{\textbf{$w_{min}$}} \\
		\hline
		\hline
		CHSH inequality (\ref{chsh}) &  $0.89$ \\
		\hline
		Chained $3$-settings Bell-CHSH inequality (\ref{chain3}) & $0.97$  \\
		\hline
		$4$-settings Gisin  inequality (\ref{gisin4}) & $0.96$  \\
		\hline
		DZC  inequality (\ref{deng4}) & $0.95$  \\
		\hline
		BG inequality (\ref{bru4}) & $0.96$  \\
		\hline
		1st AIIG inequality (\ref{avis1}) & $0.93$  \\
		\hline
		2nd AIIG inequality (\ref{avis2}) & $0.96$  \\
		\hline
	\end{tabular}
	\caption{$w_{min}$ for CHSH inequality (\ref{chsh}), chained $3$-settings Bell-CHSH inequality (\ref{chain3}), $4$-settings Gisin inequality (\ref{gisin4}), DZC  inequality (\ref{deng4}), BG inequality (\ref{bru4}), 1st AIIG inequality (\ref{avis1}), 2nd AIIG inequality (\ref{avis2}).}
	\label{tab4}
\end{table}

From Table \ref{tab4} we observe that CHSH inequality is the most robust against mixedness of the shared state in the context of sharing of bipartite nonlocality by multiple Bobs with single Alice in Scenario \ref{scenario1}. In other words, there is range of $w \in [0.89, 0.93)$  of the initial shared state for which two Bobs can sequentially demonstrate bipartite nonlocality with single Alice in Scenario \ref{scenario1} using CHSH inequality (\ref{chsh}), but not using the other local-realist inequalities considered.

Tables \ref{tab1} and \ref{tab2} reflects inequivalence of CHSH inequality (\ref{chsh}), chained $3$-settings Bell-CHSH inequality (\ref{chain3}), $4$-settings Gisin inequality (\ref{gisin4}), DZC  inequality (\ref{deng4}), BG inequality (\ref{bru4}), 1st AIIG inequality (\ref{avis1}), 2nd AIIG inequality (\ref{avis2}) with the $4$-settings Bell-CHSH inequality (\ref{chain4}), $3$-settings Gisin inequality (\ref{gisin3}), $I_{3322}$ inequality (\ref{i3322}) in the context of sharing of bipartite nonlocality in Scenario \ref{scenario1}. Because at most two Bobs can demonstrate bipartite nonlocality in Scenario \ref{scenario1} using CHSH inequality (\ref{chsh}), chained $3$-settings Bell-CHSH inequality (\ref{chain3}), $4$-settings Gisin inequality (\ref{gisin4}), DZC  inequality (\ref{deng4}), BG inequality (\ref{bru4}), 1st AIIG inequality (\ref{avis1}), 2nd AIIG inequality (\ref{avis2}). On the other hand, at most one Bob can demonstrate bipartite nonlocality in Scenario \ref{scenario1} using $4$-settings Bell-CHSH inequality (\ref{chain4}), $3$-settings Gisin inequality (\ref{gisin3}) and $I_{3322}$ inequality (\ref{i3322}). However, tables \ref{tab1} and \ref{tab2} do not demonstrate inequivalence between CHSH inequality (\ref{chsh}), chained $3$-settings Bell-CHSH inequality (\ref{chain3}), $4$-settings Gisin inequality (\ref{gisin4}), DZC  inequality (\ref{deng4}), BG inequality (\ref{bru4}), 1st AIIG inequality (\ref{avis1}), 2nd AIIG inequality (\ref{avis2}). But the interesting results revealed in tables \ref{tab3} and \ref{tab4} present the inequivalence between CHSH inequality (\ref{chsh}), chained $3$-settings Bell-CHSH inequality (\ref{chain3}), $4$-settings Gisin inequality (\ref{gisin4}), DZC  inequality (\ref{deng4}), BG inequality (\ref{bru4}), 1st AIIG inequality (\ref{avis1}), 2nd AIIG inequality (\ref{avis2}) in terms of their robustness against entanglement or mixedness of the initial shared state in the context of sharing of bipartite nonlocality in Scenario \ref{scenario1}.

\section{CONCLUSIONS} \label{sec8}

In the present study, we investigate a scenario in which a single Alice and multiple Bobs share an entangled system of two spatially separated spin-$\frac{1}{2}$ particles. Alice measures on the first particle and multiple Bobs measure on the second particle sequentially. In this scenario it was earlier shown theoretically \cite{sygp, mal} and then experimentally \cite{exp1, exp2} that at most two Bobs can demonstrate bipartite nonlocality through the quantum violations of CHSH inequality, i. e., when each party performs two dichotomic measurements. In the present study, we address the question whether the number of Bobs who can sequentially demonstrate bipartite nonlocality with single Alice can be altered by increasing the number of dichotomic measurements performed by each observer. Interestingly, we have found that the number of Bobs cannot be increased by increasing the number of dichotomic measurements performed by each observer, contrary to the case of EPR steering \cite{sas}. In other words, we show that no local realist inequality provides advantage over CHSH inequality in the context of bipartite nonlocality sharing in the above scenario.

Till date all studies related to sharing of quantum nonlocality \cite{sygp, mal, exp1,exp2} or steering \cite{sas} consider maximally entangled pure state to be initially shared between Alice and multiple Bobs. However, preparing maximally entangled pure state in real experimental situations is quite difficult. Hence, it is legitimate to ask what will happen in the context of bipartite nonlocality sharing when a non-maximally entangled pure state or a mixed state is initially shared between Alice and multiple Bobs. In order to address this issue, we have investigated the robustness of different local realist inequalities against the entanglement as well as the mixedness of the initially shared state in the context of bipartite nonlocality sharing. The significant result revealed by the present study is that the CHSH inequality is the most robust in the case of non-maximally entangled states as well as  mixed states.

To summarize, the present study incorporates several hitherto unexplored features of quantum nonlocality sharing by multiple observers. Since it is already established that nonlocal correlations acts as resources for various quantum information processing tasks, such as device independent randomness generation \cite{appran}, key distribution \cite{appkd1, appkd2}, reductions of communication complexity \cite{appcc} etc., using a single quantum nonlocal resource many times sequentially may have practical implications in addition to its foundational significance. The present study, therefore, may have several informational theoretic applications which would be worth pursuing in future.

Investigating whether there exists any local realist inequality which provides advantage over CHSH inequality in the context of bipartite nonlocality sharing in the above scenario is another direction for future studies. It is important to explore the concept of nonlocality sharing for higher dimensional systems by using different multi-outcome local realist inequalities \cite{cglmp, gwi}. Another area for future studies is to investigate the concept of sharing of genuine multipartite quantum nonlocality \cite{SI} or genuine multipartite quantum steering \cite{stm1, stm2, stm3} by multiple observers measuring sequentially on the same particle.

\section{ ACKNOWLEDGEMENTS}
D. D. acknowledges the financial support from University Grants Commission (UGC), Government of India. S. S. acknowledges the financial support from INSPIRE programme, Department of Science and Technology (DST), Government of India.

\end{document}